\documentclass{article}

\usepackage{arxiv}

\usepackage{multirow}
\usepackage{array}
\usepackage[ruled,vlined]{algorithm2e}

\usepackage{xcolor}

\usepackage[utf8]{inputenc} % allow utf-8 input
\usepackage[T1]{fontenc}    % use 8-bit T1 fonts
\usepackage{hyperref} 
\usepackage{scalerel,stackengine}
\stackMath
\newcommand\reallywidehat[1]{%
\savestack{\tmpbox}{\stretchto{%
  \scaleto{%
    \scalerel*[\widthof{\ensuremath{#1}}]{\kern-.6pt\bigwedge\kern-.6pt}%
    {\rule[-\textheight/2]{1ex}{\textheight}}%WIDTH-LIMITED BIG WEDGE
  }{\textheight}% 
}{0.5ex}}%
\stackon[1pt]{#1}{\tmpbox}%
}

\usepackage{graphicx}
\usepackage[export]{adjustbox}
\usepackage{xcolor}
\newcommand\Tstrut{\rule{0pt}{2.1ex}}       % "top" strut
\usepackage{subfigure}
\usepackage{colortbl}
\usepackage{paralist}
\usepackage{bbm}
\usepackage{url}            % simple URL typesetting
\usepackage{booktabs}       % professional-quality tables
\usepackage{amsfonts}       % blackboard math symbols
\usepackage{nicefrac}       % compact symbols for 1/2, etc.
\usepackage{microtype}      % microtypography
\usepackage{lipsum}	
\usepackage{xcolor}
\usepackage{xcolor}
\usepackage{paralist}
\usepackage{subfigure}
\usepackage{latexsym}
\usepackage{enumitem}
\usepackage{pgf-pie}
\usepackage{amsmath}
\usepackage{graphicx}
\usepackage[export]{adjustbox}
\usepackage{xcolor}
\usepackage{subfigure}
\usepackage{colortbl}
\usepackage{paralist}
\usepackage{bbm}
\usepackage{url}            % simple URL typesetting
\usepackage{booktabs}       % professional-quality tables
\usepackage{amsfonts}       % blackboard math symbols
\usepackage{nicefrac}       % compact symbols for 1/2, etc.
\usepackage{microtype}      % microtypography
\usepackage{lipsum}	
\usepackage{xcolor}
\title{Gender Bias, Social Bias and Representation: 70 Years of B$^H$ollywood}

\author{
Kunal Khadilkar\thanks{Kunal Khadilkar and Ashiqur R. KhudaBukhsh are equal contribution first authors.} \\
  \small{Carnegie Mellon University} \\
  \texttt{kkhadilk@cs.cmu.edu} \\
  \And
  Ashiqur R. KhudaBukhsh$^*$\\
  \small{Carnegie Mellon University}\\
  \texttt{akhudabu@cs.cmu.edu} \\
 \And
Tom M. Mitchell \\
  \small{Carnegie Mellon University}\\
  \texttt{tom.mitchell@cs.cmu.edu} \\
}

\begin{document}
\maketitle

\begin{abstract}
  With an outreach in more than 90 countries, a market share of 2.1 billion dollars and a target audience base of at least 1.2 billion people, Bollywood, aka the Mumbai film industry, is a formidable entertainment force. While the number of lives Bollywood can potentially touch is massive, no comprehensive NLP study on the evolution of social and gender biases in Bollywood dialogues exists. Via a substantial corpus of movie dialogues spanning a time horizon of 70 years, we seek to understand the portrayal of women, in a broader context studying subtle social signals, and analyze the evolving trends in geographic and religious representation in India. Our argument is simple -- popular movie content reflects social norms and beliefs in some form or shape. In this project, we propose to analyze such trends over 70 years of Bollywood movies contrasting them with their Hollywood counterpart and critically acclaimed world movies.
\end{abstract}

% keywords can be removed

\section{Introduction}
\noindent \emph{What types of social biases can we analyze and detect through the lens of a diachronic corpus of popular entertainment?} In this paper, we focus on Bollywood, aka the Mumbai film industry, and analyze a curated corpus of film subtitles for the last 70 years. While Bollywood is an entertainment industry worth billions and has a target audience of 1.2 billion people, little or no work exists that analyzed a wide range of social biases and signals that can be uncovered from this rich corpus. In this work, we contrast our findings with an analogous corpus of Hollywood and for a specific subset of research questions, we dig deeper and look into world movies.       

\begin{table*}[h!]
\begin{center}
     \begin{tabular}{| p{6cm}  | p{6cm} |}
    \hline
     \cellcolor{blue!25} Akeli ladki khuli tijori ki tarah hoti hai (Jab We Met) &\cellcolor{red!25} A girl who is alone is like an open treasure. (Jab We Met) \\
     \hline 
Generated Movie Revenue: $\approx$\$14,899,137
 &  \\
    \cellcolor{blue!25} Marriage se pehle ladkiyajn sex object hoti hain, our marriage ke baad they object to sex! (Kambakkht Ishq) &\cellcolor{red!25} Before marriage, girls are sex objects, and after marriage, girls object to sex.  (Kambakkht Ishq) \\
    \hline
Generated Movie Revenue: $\approx$\$17,531,586
 &  \\
\cellcolor{blue!25} Tu ladki ke peeche bhagega, ladki paise ke peeche bhagegi. Tu paise ke piche bhagega, ladki tere peeche bhagegi (Wanted) &\cellcolor{red!25} You are chasing the girls, while the girls are chasing money. If you start chasing money, girls will automatically chase you. (Wanted) \\
    \hline
Generated Movie Revenue: $\approx$\$27,630,059
 &  \\ \hline

    \end{tabular}
    % \vspace{0.2cm}
\end{center}
    \caption{Illustrative examples of misogynistic dialogues present in blockbuster Bollywood movies (movie names are presented in parentheses).  The dialogues (in blue) are in Romanized Hindi, and their approximate English translations are highlighted in red. }
\label{tab:dialogexamples}
%\vspace{-.2in}
\end{table*}

Our primary focus is gender bias. As shown in Table~\ref{tab:dialogexamples}, several Bollywood movies are riddled with sexist and misogynist dialogues. It is thus not surprising that cutting edge NLP methods would reveal some of these existing biases. We are, however, interested in a broader research question: \emph{In a developing nation, what kind of social insights can be gleaned from popular entertainment?} Is it possible to understand subtle gender biases such as son's preference? Can we track the evolving nature of retrograde social practises like dowry? 

Our second focus in this work is broader representation questions such as geographic representation, religious representation and caste representation. In our mixed method analyses, we identify that (1) some of gender biases observed in Bollywood is very much present in it's Western counterpart; (2) a positive trend is witnessed in observing reduced biases with progress of time and; and (3) a similar trend is observed in religious and geographic representation, with a considerable scope for improved diversity and inclusion.  

%In this paper, we analyze a curated corpus of film subtitles from the Bollywood and Hollywood film industry, spread over 70 years (1950 to 2020) and present our preliminary findings on three aspects (1) social bias toward a fair skin color (2) gender biases, and (3) gender representation. While studies analyzing gender stereotypes across different languages \cite{lewis2020gender} and detecting bias in word embeddings \cite{garg2018word} exist, barring few lines of work focusing on a single Bollywood movie~\cite{chatterjee2016english} or a small subset of movies~\cite{khan2018gender}, no existing work has focused on analyzing social biases and under-representation in Bollywood films. We perform a first ever large-scale study comprising 1400 films, to uncover implicit social and racial biases in the entertainment industry, using computational science and natural language processing techniques. 

%In Figure~\ref{fig:wordCloud}, what do the prominent presence of negative adjectives such as \texttt{wanton} in old Bollywood movies (see, Figure~\ref{fig:oldBolly}) juxtaposed with the presence of positive verbs such as \texttt{respect} (see, Figure~\ref{fig:newBolly}) tell us? In this paper, we explore a wide array of NLP  techniques to analyze our research questions through the lens of popular movies. We use

\section{Research questions}
We aim to answer the following research questions - \\ 
\emph{\textbf{RQ 1:} How is gender bias reflected through movie dialogues in the Bollywood and Hollywood movie industries?} \\ 
\emph{\textbf{RQ 2:} How do award winning foreign feature films compare with Bollywood and Hollywood in addressing gender equality?
Does genre make any difference?} \\
\emph{\textbf{RQ 3}: Is beauty associated with fair skin in the movie dialogues describing women?} \\
\emph{\textbf{RQ 4:} Does Bollywood reflect the well-documented son's preference in medical and social science research? How has the sentiment around retrograde social practices such as dowry evolved?} \\
\emph{\textbf{RQ 5:} Which geographical areas have been consistently underrepresented in the Indian film industry?} \\
\emph{\textbf{RQ 6:} Can we gain an insight into the religious representation of a country, through a film corpus spanning 70 years?} \\
\emph{\textbf{RQ 7:} Can we extract economic signals through popular film dialogues?}  \\
\emph{\textbf{RQ 8:} How are religions perceived in movies? Can we track evolving national priorities from popular entertainment?}

\section{Related Work}
% The Bollywood, or the Indian film industry has become a powerhouse in the 20th century, with a valuation of 11 Billion rupees and an audience base of more than 1.2 Billion people. Indian films are showcased in more than 90 countries over the world, at various international film festivals. With such a growing influence on the entertainment industry around the world, it becomes necessary to study the biases that are showcased over the years through the cinematographic medium. 

%Beyond blog articles \cite{blog1,blog2,blog3} and news articles or social media backlashes on platforms like Twitter, little or no NLP analysis exists on Bollywood. 
Studies analyzing gender stereotypes across different languages \cite{lewis2020gender} and detecting bias in word embeddings \cite{garg2018word} use books and news data sources for their analyses. Existing lines of work in uncovering entertainment industry bias focus on a single Bollywood movie~\cite{chatterjee2016english} or a small subset of movies~\cite{khan2018gender}. \cite{bmovies} focused on plot points and film information taken from Wikipedia. We consider a different and potentially richer data set of film subtitles spanning 70 years. We contrast our work with Hollywood and award winning world movies, and our analyses cover a broader set of aspects such as retrograde social practices, uncovering subtler biases and highlighting geographical and religious underrepresentation. Unlike previous work on movie subtitles~\cite{lewis2020gender}, our focus is on Bollywood content largely ignored by the information science research community so far. 

In terms of our dataset, our work is most similar to~\cite{khadilkarBollywood} that looked at a smaller corpus of Bollywood subtitles.  Our work contrasts with~\cite{khadilkarBollywood} in the following key ways. First, our treatment to the analyses of gender bias  (1) includes diachronic-word embedding analysis and word embedding association tests (WEAT); (2) is grounded on well-established lexicons; (3) looks into subtler signals such as son's preference; (4) tracks retrograde social practices such as dowry and (5) considers additional data sources (e.g., world movies). Second, our work tackles important  questions on geographic and religious representation unaddressed by~\cite{khadilkarBollywood}. Finally, we look into questions related to economic signals and show promising computational creativity results. 

%No existing work has focused on analyzing social biases and under-representation in Bollywood films on a diachronic dialogue corpus. We perform a first ever large-scale study comprising more than 2000 films' dialogues, to uncover implicit social and racial biases in the entertainment industry. \\

% Our work is related to \cite{ECAIElection},~\cite{petroni-etal-2019-language},\cite{hamilton-etal-2016-diachronic},\cite{weatsource} in a sense that we employ this broad suite of NLP techniques on a novel domain of popular entertainment. 
In this paper, we explore a wide array of NLP  techniques to analyze our research questions through the lens of popular movies. We use
\begin{compactenum}
\item Simple \emph{count-based statistics} relying on highly popular lexicons~\cite{warriner2013norms} and gender representation studies~\cite{twenge2012male,senden2015she};
\item \emph{Cloze test}, an analysis technique that have a solid grounding in psycholinguistics literature~\cite{taylor1953cloze,smith2011cloze}. To the best of our knowledge, for the first time, we explore a recent technique~\cite{petroni-etal-2019-language} previously used to mine political insights~\cite{ECAIElection} in the context of uncovering social biases. Through a series of cloze tests on a language model~\cite{devlin2018bert} fine-tuned on our data sets, we present our  findings.
\item \emph{Analysis of aligned diachronic word embedding spaces} using recently proposed techniques~\cite{hamilton-etal-2016-diachronic}.
\item \emph{Free form text completion using GPT-2 }\cite{gpt2}, for a novel task of tracking economic signals.  
\end{compactenum}
We employ this broad suite of NLP techniques on a novel domain of popular entertainment.\\ 

Geographical and community representation in India has been studied by various political and social scientists. \cite{mukherjee} showcase the population misrepresentation in political settings, while \cite{haoginlen} focuses on underrepresentation of north-east India in mainstream newspapers. Our work complements this research \cite{haoginlen} and presents corroborating evidence from a very different data source.

% Our work draws inspiration from social science research to track the trajectory of Indian development through the years. 

% \begin{itemize}
%     \item Gender bias analysis in Bollywood (one FACCT paper mentioned by Harvard CRCS reviewer and we added that to AAAI). They focused on plot points taken from wikipedia - we look at a much richer data set of film subtitles. We contrast our work with Hollywood and award winning world movies. Our analyses cover a broader set of aspects such as retrograde social practices,   
%     \item Lupyan et al. focused film subtitles, but our focus is on Bollywood content largely ignored by information science research community so far. 
%     \item Our work is related to [all the methods] in a sense that we emply this broad suite of NLP techniques on a novel domain. 
%     \item Our work draws inspiration from social science research several social aspects tracking the trajectory of Indian development through the years.  
% \end{itemize}

\section{Data Set} 
We construct the following two data sets of movie subtitles.\\
\textbf{1. Bollywood movies, $\mathcal{D}_\mathit{bolly}$:} We consider movies spanning seven decades (1950--2020). For each decade, we retrieved subtitles \cite{lison2016opensubtitles2016} of 100 movies (700 total films). \newline 
\textbf{2. Hollywood movies, $\mathcal{D}_\mathit{holly}$:} Similar to Bollywood movies, we considered 100 top-grossing movies from each of the seven decades (700 total films). Overall, $\mathcal{D}_\mathit{bolly}$ and $\mathcal{D}_\mathit{holly}$ consist of 1.1M dialogues (6.2M Tokens) and 1M dialogues (5.4M Tokens), respectively.

For a subset of our analyses, we divide our corpus into three buckets presented in Table~\ref{tab:dataset}.

% \begin{table}[]
% \begin{tabular}{|l|l|l|}
% \hline
% \textbf{Corpus} & \textbf{Industry} & \textbf{Time Period} \\ \hline
% Dbolly\_old     & Bollywood         & 1950 - 1969          \\ \hline
% Dbolly\_mid     & Bollywood         & 1970 - 1999          \\ \hline
% Dbolly\_new     & Bollywood         & 2000 - 2020          \\ \hline
% \end{tabular}
% \end{table}

\begin{table}[h]
\centering
\begin{tabular}{ll}
\begin{tabular}{|l|l|l|}
\hline
\Tstrut 
Corpus & Industry & Time Period \\ \hline
\Tstrut
$\mathcal{D}_\mathit{bolly}^\mathit{old}$    & Bollywood         & 1950 - 1969          \\ \hline
\Tstrut
$\mathcal{D}_\mathit{bolly}^\mathit{mid}$     & Bollywood         & 1970 - 1999          \\ \hline
\Tstrut
$\mathcal{D}_\mathit{bolly}^\mathit{new}$    & Bollywood         & 2000 - 2020          \\ \hline

\end{tabular}
&
\begin{tabular}{|l|l|l|}
\hline
\Tstrut 
Corpus & Industry & Time Period \\ \hline
\Tstrut
$\mathcal{D}_\mathit{holly}^\mathit{old}$    & Hollywood         & 1950 - 1969          \\ \hline
\Tstrut
$\mathcal{D}_\mathit{holly}^\mathit{mid}$     & Hollywood         & 1970 - 1999          \\ \hline
\Tstrut 
$\mathcal{D}_\mathit{holly}^\mathit{new}$    & Hollywood         & 2000 - 2020          \\ \hline
\end{tabular}
\end{tabular}
\vspace{0.5cm}
    \caption{Dataset split for Bollywood and Hollywood}
    \label{tab:dataset}
\end{table}
We further collect 150 movies which have been nominated for the \emph{Best International Feature Film} award 1970 onwards, at the Oscars (\url{https://www.oscars.org}) for a subset of our analyses. 

\section{Gender Bias in Movie Dialogues} 

\emph{\textbf{RQ 1:} How is gender bias reflected through movie dialogues in the Bollywood and Hollywood movie industries?} \\

\subsection{Gendered pronouns}

\begin{figure}[h!]
\centerline{\includegraphics[totalheight=4cm,width=6cm ]{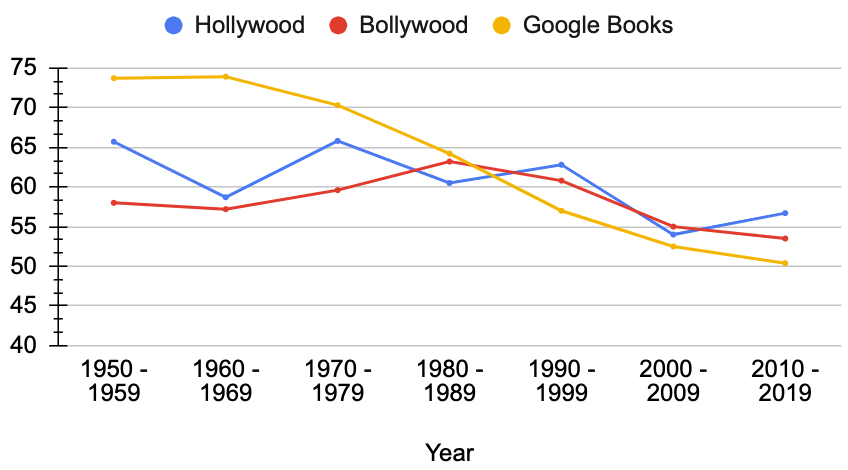}}
\vspace{-0.3cm}
    \caption{$\mathit{MPR}$ in $\mathcal{D}_\mathit{bolly}$ and $\mathcal{D}_\mathit{holly}$ }
    \label{fig:verticalcell}
\end{figure}

Following extensive literature on gendered pronouns' relative distributions and their implications~\cite{twenge2012male, senden2015she}, we consider a simple measure of gender representation:  relative occurrence of pronouns of each gender (Men: he, him. Women: she, her). Let $\mathcal{N}_\mathit{w}$ denote the number of times a token $w$ appears in a corpus. We define  \emph{Male Pronoun Ratio (MPR)} as follows: 
    $\mathit{MPR} = \frac{\mathcal{N}_\mathit{he} + \mathcal{N}_\mathit{him} }{\mathcal{N}_\mathit{he} + \mathcal{N}_\mathit{him}+ \mathcal{N}_\mathit{she} + \mathcal{N}_\mathit{her}}*100
$.
Figure~\ref{fig:verticalcell} plots \emph{MPR} of our decade-wise movie data sets and contrasts with \emph{MPR} computed using google n-grams. Our results indicate that even now, both Bollywood and Hollywood exhibit comparable skew in gendered pronoun usage.

\subsection{Cloze tests}

When presented with a sentence (or a sentence stem) with a missing word, a cloze task~\cite{taylor1953cloze} is essentially a fill-in-the-blank task. For instance, in the following cloze task: \emph{In the} \texttt{[MASK]}\emph{, it is very sunny}, \texttt{summer} is a likely completion for the missing word. Given a cloze test, \texttt{BERT}, a well-known language model~\cite{devlin2018bert}, outputs a series of token ranked by probability. In fact, in the above cloze test, the top three tokens (ranked by probability) predicted by \texttt{BERT}$_\mathit{base}$ are: summer, winter and spring. Recent lines of research has explored \texttt{BERT}'s masked query prediction for (1) knowledge base extraction~\cite{petroni-etal-2019-language} and (2) mining political insights~\cite{ECAIElection}.     

\begin{table*}[h]
\scriptsize
\begin{center}
     \begin{tabular}{| p{1cm}  | p{2.1cm} | p{2.1cm} | p{2.1cm} | p{2.1cm} | p{2.1cm} |}
    \hline
\textbf{Probe} & \textbf{\texttt{BERT}\textsubscript{base}}                                                               & \textbf{\texttt{BERT}\textsubscript{$\mathcal{D}_\mathit{bolly}^\mathit{old}$}}                                                & \textbf{\texttt{BERT}\textsubscript{$\mathcal{D}_\mathit{bolly}^\mathit{new}$}}                                                       & \textbf{\texttt{BERT}\textsubscript{$\mathcal{D}_\mathit{holly}^\mathit{old}$}}                                                            & \textbf{\texttt{BERT}\textsubscript{$\mathcal{D}_\mathit{holly}^\mathit{new}$}}                                                              \\ \hline
$\mathit{cloze}_1$         & man, widow, woman, \textcolor{blue}{doctor}, \textcolor{red}{slave}, \textcolor{blue}{soldier}, bachelor, \textcolor{blue}{merchant}, \textcolor{blue}{farmer}, \textcolor{blue}{lawyer}, \textcolor{red}{servant} [\textcolor{black}{\textbf{4.8}}] & \textcolor{red}{prostitute}, \textcolor{red}{servant}, woman, \textcolor{red}{slave}, bachelor, \textcolor{blue}{doctor}, \textcolor{blue}{lawyer}, man, widow, \textcolor{red}{maid}, \textcolor{blue}{worker} [\textcolor{black}{\textbf{4.64}}] & \textcolor{blue}{doctor}, woman, \textcolor{red}{servant}, \textcolor{blue}{lawyer}, \textcolor{red}{maid}, hindu, \textcolor{blue}{nurse}, \textcolor{blue}{teacher}, \textcolor{blue}{gardener}, lady, man [\textcolor{black}{\textbf{5.7}}] & woman, \textcolor{red}{slave}, \textcolor{red}{servant}, \textcolor{blue}{nurse}, lady, man, \textcolor{blue}{teacher}, \textcolor{blue}{lawyer}, peasant, \textcolor{red}{maid}, wife [\textcolor{black}{\textbf{5.3}}] & woman, \textcolor{blue}{lawyer}, \textcolor{blue}{doctor}, \textcolor{blue}{nurse}, \textcolor{blue}{teacher}, man, \textcolor{blue}{writer}, \textcolor{blue}{secretary}, \textcolor{red}{prostitute}, professional, \textcolor{blue}{carpenter} [\textcolor{black}{\textbf{5.7}}] \\ \hline
$\mathit{cloze}_2$         & man, \textcolor{blue}{soldier}, gentleman, \textcolor{blue}{farmer}, \textcolor{blue}{merchant}, woman, \textcolor{red}{slave}, bachelor, \textcolor{blue}{doctor}, \textcolor{blue}{carpenter}, \textcolor{red}{servant} [\textcolor{black}{\textbf{5.48}}] & man, \textcolor{blue}{gentleman}, \textcolor{blue}{lawyer}, lawyer, \textcolor{red}{servant}, \textcolor{blue}{doctor}, \textcolor{blue}{farmer}, \textcolor{blue}{worker}, \textcolor{blue}{craftsman}, \textcolor{red}{slave}, \textcolor{red}{criminal} [\textcolor{black}{\textbf{5.0}}] & \textcolor{blue}{doctor}, \textcolor{blue}{lawyer}, \textcolor{blue}{policeman}, man, \textcolor{blue}{farmer}, bachelor, \textcolor{blue}{gardener}, \textcolor{red}{servant}, \textcolor{blue}{soldier}, \textcolor{blue}{mechanic}, \textcolor{blue}{builder} [\textcolor{black}{\textbf{5.3}}] & \textcolor{blue}{carpenter}, \textcolor{blue}{policeman}, \textcolor{blue}{lawyer}, \textcolor{blue}{soldier}, \textcolor{blue}{farmer}, \textcolor{blue}{gentleman}, \textcolor{red}{servant}, man, peasant, \textcolor{red}{slave}, \textcolor{blue}{doctor} [\textcolor{black}{\textbf{5.0}}] & man, \textcolor{blue}{lawyer}, \textcolor{blue}{soldier}, \textcolor{blue}{doctor}, \textcolor{blue}{carpenter}, \textcolor{blue}{gentleman}, \textcolor{blue}{clergyman}, \textcolor{blue}{farmer}, \textcolor{blue}{writer}, \textcolor{blue}{craftsman}, \textcolor{blue}{minister} [\textcolor{black}{\textbf{5.78}}] \\  \hline
    \end{tabular}
\end{center}
\vspace{0.2cm}
\caption{{Cloze test results. Predicted tokens are ranked by decreasing probability. Positive and negative words are color coded with blue and red, respectively. The number in the bracket represents the average valence score (obtained from~\protect\cite{warriner2013norms}) calculated for the answers to the cloze test.}}
\label{tab:allCloze1}
%%\vspace{-.2in}
\end{table*}

Our cloze test results are summarized in Table~\ref{tab:allCloze1}. We observe  that completion results for both genders across both movie industries improve over time. In order to quantify the completion results, we consider a well-known lexicon of emotional valence ratings~\cite{warriner2013norms} of nearly 14,000 English words to quantify the change of cloze test completions over time. The valence score of these words are presented in a scale of 1 to 10 with 10 indicating  highly positive and 1 indicating highly negative. For example, the emotional valence score of \texttt{happy} and \texttt{sad} are 8.47 and 2.10, respectively. For a given data set and a cloze test pair, we compute the average valence score of the completions (listed in square brackets in Table~\ref{tab:allCloze}). We further note that comparing completion words across the two genders can be technically difficult simply because we observe that there can be potential bias in the scores. For instance, the valence scores for \texttt{man} and \texttt{woman} are 5.42 and 7.09, respectively. Hence we restrict ourselves to comparing within a specific gender for a given movie industry.    

\begin{table}[htb]
\centering
\begin{tabular}{|l|l|l|}

\hline
                & \textbf{Bollywood} & \textbf{Hollywood} \\ \hline
Women & 22\%               & 7\%                \\ \hline
Men & 5\%                & 15\%               \\ \hline
\end{tabular}
\vspace{0.2cm}
\caption{Percent increase in average valence score for cloze test completions between old movies and new movies.}
\label{tab:increase}
\end{table}

Table~\ref{tab:increase} lists the percentage of increase in the valence score of the completion for a particular gender across different movie industries. We note that for both Bollywood and Hollywood, the valence scores for both genders improved over time. However, for Bollywood, we notice that rate of increase for women is substantially more pronounced than that for men. This observation aligns with the continual fight for gender equality in India~\cite{nayak2012women} and major movements that have mobilised voices for women's right to work~\cite{chen1995matter}, financial independence~\cite{goyal2011women}, and marital laws~\cite{nigam2005understanding}.

\subsection{Diachronic word embeddings}

\begin{figure}[htb]
% \vspace{-.15cm}
\centering
\subfigure[\texttt{Woman} over the years]{%
\includegraphics[frame,width = 0.48 \textwidth, height=0.30\textwidth]{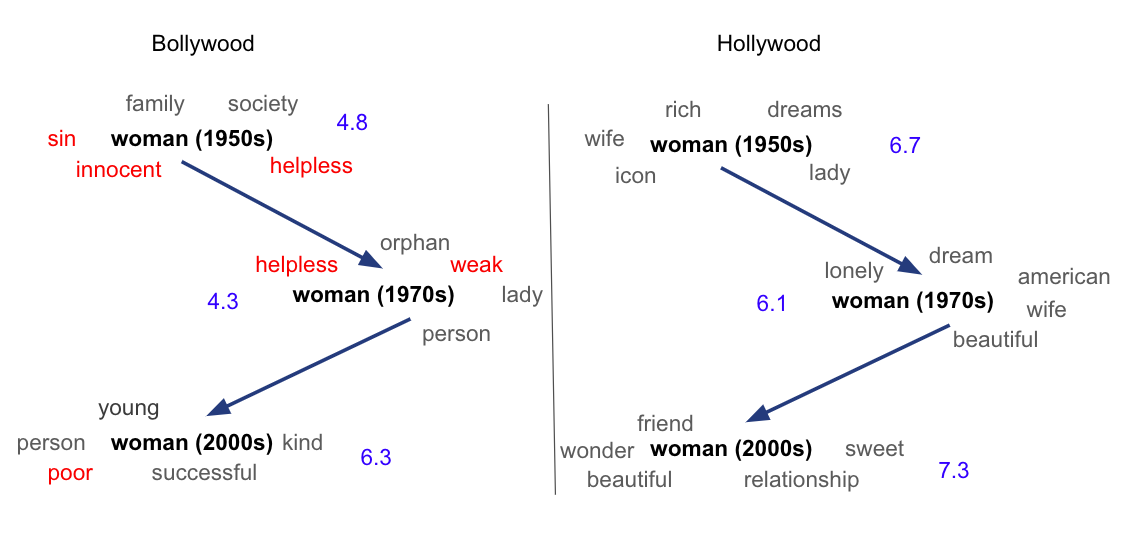}
\label{fig:oldBolly}}
\subfigure[\texttt{Man} over the years]{%
\includegraphics[frame,width = 0.48 \textwidth, height=0.30\textwidth]{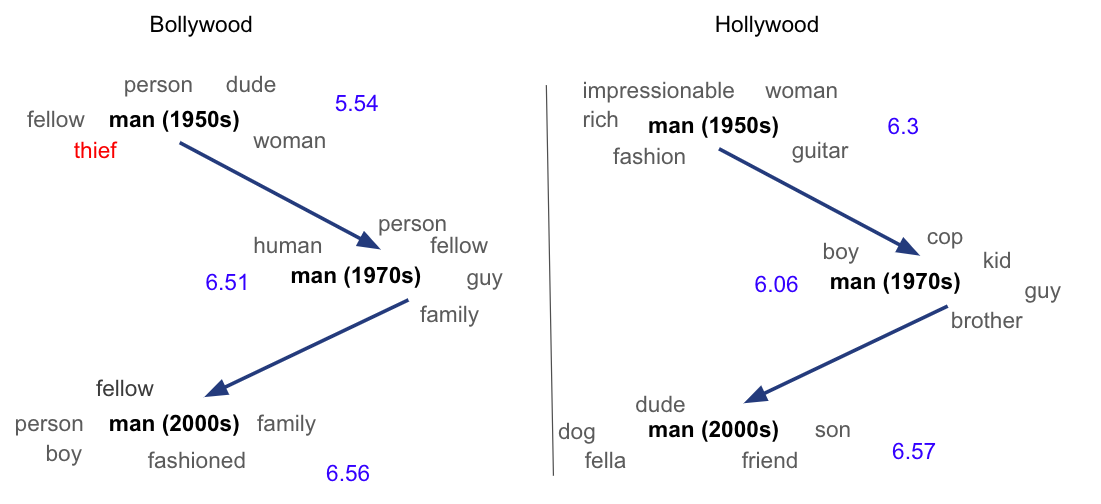}
 \label{fig:newBolly}}
\caption{Nearest neighbors of \texttt{man} and \texttt{woman} over the years. The overall average valence of nearest neighbors according to the lexicon provided in~\protect\cite{warriner2013norms} for a given time-period is presented in blue font.}
\end{figure}

The meaning of words and the context in which they are used change over time \cite{xie-etal-2019-text}. The language spoken in a community is representative of the cultural norms and customs followed in that region. Existing hypotheses \cite{bookbybee} indicate that word frequency may play a role in changing the meaning of the words over time. The meanings of less-frequent words are more susceptible to drifts than that of highly frequent words.  \cite{hamilton-etal-2016-diachronic} provide a robust multilingual approach to align diachronic word embeddings using orthogonal procrustes. We follow the same method to align different sub-corpora for Bollywood and Hollywood. 
In addition to the four sub-corpora --  $\mathcal{D}_\mathit{bolly}^\mathit{old}$,  $\mathcal{D}_\mathit{bolly}^\mathit{new}$, $\mathcal{D}_\mathit{holly}^\mathit{old}$,  $\mathcal{D}_\mathit{holly}^\mathit{new}$ -- we consider the time period of 1970-2000 and construct two additional sub-corpora $\mathcal{D}_\mathit{bolly}^\mathit{mid}$ and $\mathcal{D}_\mathit{holly}^\mathit{mid}$.We train word2vec \cite{NIPS2013_5021} with SGNS (Skip-gram with Negative Sampling), to create embeddings for each of the buckets mentioned above. Let $\mathcal{W}^{(t)}\in \mathbb{R}^{(d)\times \mathcal{V}}$ be the matrix of word embeddings learnt for period $t$ for vocabulary $\mathcal{V}$. Following \cite{hamilton-etal-2016-diachronic}, we align the word embeddings using the top 10,000 common tokens present across time periods $t$ and $t+1$ by optimizing:
\begin{equation}
    R^{t} = arg \min_{Q^{T}Q=I}\left \| Q\mathcal{W}^{t} - \mathcal{W}^{(t+1)} \right \|_{F} 
\end{equation}

where $R^{t} \in \mathbb{R}^{d \times d}$.

We focus on the portrayal of women and men using these aligned embeddings. We observe that the valence scores for both genders across both movie industries show a similar pattern. The scores are the lowest during the 1970-1999 time period. The valence scores for the newer movies are better than the scores for the older movies. The dip in the valence scores during the time-period of 1970-1999 in India can be ascribed to a social and cultural crisis influenced by an unstable political climate (assassinations of two prime ministers~\cite{hardgrave1985india,kaarthikenyan2008rajiv}), two major wars between India and Pakistan \cite{kashmirbook,kashmirbook2}, and a large overlap with a pre-economic liberalization period~\cite{pedersen2000explaining}. \\    

\subsection{Word Embedding Association Test (WEAT) for quantifying gender bias}
While valence scores give us an indication regarding the difference in valence (degree of pleasantness) scores between the two genders on Language Models cloze tests, we explore quantifying the bias in our corpus using a more popular technique of Word Embedding Association Test (WEAT) \cite{weatsource}. For the purpose of this analysis, We train the sub-corpora using GloVe \cite{glove} embeddings, as given in \cite{weatsource}, to understand the evolution of gender bias through time. According to WEAT, we consider two equal sized sets of occupations, $S_{1}$ and $S_{2}$ and two sets of attribute words $A_{1}$ and $A_{2}$. 

The similarity of two words, say $x$ and $y$, is given by calculating the cosine similarity of the corresponding word embeddings, $cos(w_{x},w_{y})$. As given in \cite{weatsource}, the differential association of a word c with word sets $A_{1}$ and $A_{2}$ is given by:

\begin{equation}
    g(c,A_{1},A_{2},w) = {\overset{mean}{a\epsilon A_{1}}} cos(w_{c},w_{a}) - {\overset{mean}{b\epsilon A_{2}}} cos(w_{c},w_{b})
\end{equation}

Further, the WEAT is calculated as:
\begin{equation}
    B_{weat}(w) = \frac{{\overset{mean}{s\epsilon S_{1}}} g(s,A_{1},A_{2},w) - {\overset{mean}{t\epsilon S_{2}}} g(t,A_{1},A_{2},w)}{{\overset{std-dev}{s\epsilon S_{1} S_{2}}} g(s,A_{1},A_{2},w)}
\end{equation}

Here, the occupation sets taken from \cite{bolukbasi} $S_{1}$ and $S_{2}$ are: \\
$S_{1}$ = [``maestro'', ``skipper'', ``protege'', ``philosopher'', ``captain'', ``architect'', ``financier'', "warrior'', ``broadcaster'',``magician'', ``pilot'', ``boss''] 
$S_{2}$ = [``homemaker'', ``nurse'', ``receptionist'', ``librarian'', ``socialite'',``hairdresser'', ``nanny'', ``bookkeeper'', ``stylist'', ``housekeeper'', ``designer'', ``counselor'']

And the attribute word sets $A_{1}$ and $A_{2}$ are: 
$A_{1}$ = [`he',`man',`male'] 
$A_{2}$ = [`she',`woman',`female']

\begin{table}[h]
\centering
\begin{tabular}{|c|c|c|c|c}
\cline{1-4}
\Tstrut
Industry              & $\mathit{Old}$ & $\mathit{Mid}$ & $\mathit{New}$ &  \\ \cline{1-4}
\Tstrut
Bollywood     & 0.601        & 0.559        & 0.543        &  \\ \cline{1-4}
\Tstrut
Hollywood     & 0.456        & 0.443       & 0.410       &  \\  \cline{1-4}
\end{tabular}
\vspace{0.5cm}
\caption{Average \texttt{WEAT} scores for Bollywood and Hollywood across different time periods. A larger value indicates greater gender bias.}
\label{tab:industry_weat}
\end{table}

As shown in Table~\ref{tab:industry_weat}, we find that the average \texttt{WEAT} scores across both industries reduced over time. As compared to Bollywood, for any given time period, Hollywood exhibits lesser gender bias.

\section{Genre Specific Analysis}

\emph{\textbf{RQ 2:} How do award winning foreign feature films compare with Bollywood and Hollywood in addressing gender equality?
Does genre make any difference?} \\

Along with performing a comparative study between Bollywood and Hollywood, we are further interested in comparing the \texttt{WEAT} scores of Bollywood and Hollywood with that of a set of critically acclaimed world movies. To this end, we construct a corpus of 150 movies  nominated at the foreign film category at the Academy awards.    
%nominated foreign feature films at the Academy awards, . 

% We divide our corpus into three buckets: (1) Films from 1950 to 1969 ($\mathcal{C}_\mathit{bolly}^\mathit{old}$ and $\mathcal{C}_\mathit{holly}^\mathit{old}$); (2) Films from 1970 to 1999 ($\mathcal{C}_\mathit{bolly}^\mathit{mid}$ and $\mathcal{C}_\mathit{holly}^\mathit{mid}$);  and (3) Films from 2000 to 2019 ($\mathcal{C}_\mathit{bolly}^\mathit{new}$ and $\mathcal{C}_\mathit{holly}^\mathit{new}$). We further create $\mathcal{C}_\mathit{foreign}^\mathit{mid}$ and $\mathcal{C}_\mathit{foreign}^\mathit{new}$ which are nominated international feature films at the Academy awards, from the time period of 1970-1999 and 2000-2019 respectively. 

%Our Foreign Feature Film corpus contains 150 films in total. To cover an extensive set of Bollywood and Hollywood films, we consider multiple distinct subsets of movies (4 in total for each industry,  150 movies in each subset) and compute the average WEAT score. 

% \begin{table}[h]
% \centering
% \begin{tabular}{|c|c|c|c|c}
% \cline{1-4}
% Industry              & \textbf{Old} & \textbf{Mid} & \textbf{New} &  \\ \cline{1-4}
% Bollywood     & 0.601        & 0.559        & 0.543        &  \\ \cline{1-4}
% Hollywood     & 0.456        & 0.443       & 0.410       &  \\ \cline{1-4}
% Foreign Feature Films & --           & \textbf{0.3518}           & \textbf{0.341}        &  \\ \cline{1-4}
% \end{tabular}
% \vspace{0.5cm}
% \caption{WEAT scores for Bollywood, Hollywood and Foreign Feature Films}
% \label{tab:industry_weat}
% \end{table}

\begin{table}[h]
\centering
\begin{tabular}{|c|c|}
\cline{1-2}
\Tstrut
Industry              & \texttt{WEAT} score\\ \cline{1-2}
\Tstrut
Bollywood     & 0.523\\ \cline{1-2}
\Tstrut
Hollywood     & 0.504\\ \cline{1-2}
\Tstrut
World movies & \textbf{0.285}\\ \cline{1-2}
\end{tabular}
\vspace{0.5cm}
\caption{\texttt{WEAT} scores for Bollywood, Hollywood and world movies.}
\label{tab:bollyholly_weat}
\end{table}

%The scores for Bollywood and Hollywood are average over the 4 subsets, with Bollywood having a Standard Deviation of 0.368, and Hollywood having a Standard Deviation of 0.226

% Our results showcase that indeed over the years, the bias has decreased considerably both in Bollywood and Hollywood. A surprising result obtained are the WEAT scores for the foreign feature films. These films tend to be the least biased, both in mid and newer time periods. Achieving gender parity in dialogues, showcasing progressive or pathbreaking content in movies might be a few possible reasons for the films being nominated at the Academy awards. Note that we couldn't find relevant subtitles for foreign films in the old period (1950-1969), hence only $\mathcal{C}_\mathit{foreign}^\mathit{mid}$ and $\mathcal{C}_\mathit{foreign}^\mathit{new}$ are considered for our analysis. \\

Our results indicate that the average \texttt{WEAT} score obtained for nominated foreign feature films is the lowest as compared to the average \texttt{WEAT} score for Bollywood and Hollywood. 

%Achieving gender parity in dialogues, showcasing progressive or pathbreaking content in movies might be a few possible reasons for the films being nominated at the Academy awards. We further split the three industries into three distinct time periods (Old, Mid and New) as mentioned in the dataset section, and we see that over the years, the bias in all the three industries has decreased considerably. (Table in Appendix.)

Adventure/Action and Romance are the two most popular genres across different industries, with hundreds of films released every year. Action films generally tend to be male dominated, compared to Romantic films. We explore this hypothesis using \texttt{WEAT}, for the given genres. For this analysis, we consider four token-balanced sub corpora, $G_{bolly}^{action}$,$G_{bolly}^{romance}$,$G_{holly}^{action}$,$G_{holly}^{romance}$, each containing 150 films. All the films were post 1990, and were chosen based on the genre lists/tags given by imdb and Google. 

\begin{table}[h]
\centering
\begin{tabular}{|l|l|l|}

\hline
\Tstrut
                & $\mathit{Romance}$ & $\mathit{Action}$ \\ \hline
                \Tstrut
Hollywood & 0.079               & \textbf{0.591}                \\ \hline
\Tstrut
Bollywood & 0.343                & \textbf{0.512}               \\ \hline

\end{tabular}
\vspace{0.2cm}
\caption{\texttt{WEAT} scores for Romance and Action films.}
\label{tab:genre_weat}
\end{table}

Table \ref{tab:genre_weat} shows that the gender bias for action movies is a lot more pronounced than that in Romance movies. Across both industries and movie genres, Hollywood action films exhibit most bias.  

\section{Associating Beauty with Fair Skin}

\emph{\textbf{RQ 3}: Is beauty associated with fair skin in the movie dialogues describing women?}  \\

\begin{table*}[h]
\scriptsize
\begin{center}
     \begin{tabular}{| p{2.1cm} | p{2.1cm} | p{2.1cm} | p{2.1cm} | p{2.1cm} |}
    \hline
 \textbf{\texttt{BERT}\textsubscript{base}}                                                               & \textbf{\texttt{BERT}\textsubscript{$\mathcal{D}_\mathit{bolly}^\mathit{old}$}}                                                & \textbf{\texttt{BERT}\textsubscript{$\mathcal{D}_\mathit{bolly}^\mathit{new}$}}                                                       & \textbf{\texttt{BERT}\textsubscript{$\mathcal{D}_\mathit{holly}^\mathit{old}$}}                                                            & \textbf{\texttt{BERT}\textsubscript{$\mathcal{D}_\mathit{holly}^\mathit{new}$}}                                                              \\ \hline

soft, beautiful, pale, tanned, smooth & \textcolor{red}{fair}, no, pale, tanned, tan & \textcolor{red}{fair}, tanned, golden, smooth, pale & \textcolor{red}{fair}, pale, blue, golden, gold & \textcolor{red}{fair}, pale, tanned, golden, dark \\ \hline
    \end{tabular}
\end{center}
\vspace{0.2cm}
\caption{{Cloze test results for the probe \emph{A beautiful woman should have} \texttt{[MASK]} \emph{skin.}.}}
\label{tab:allClozeSkin}
%%\vspace{-.2in}
\end{table*}

\begin{figure}[h]
\centerline{\includegraphics[frame,totalheight=5cm,width=0.8\linewidth ]{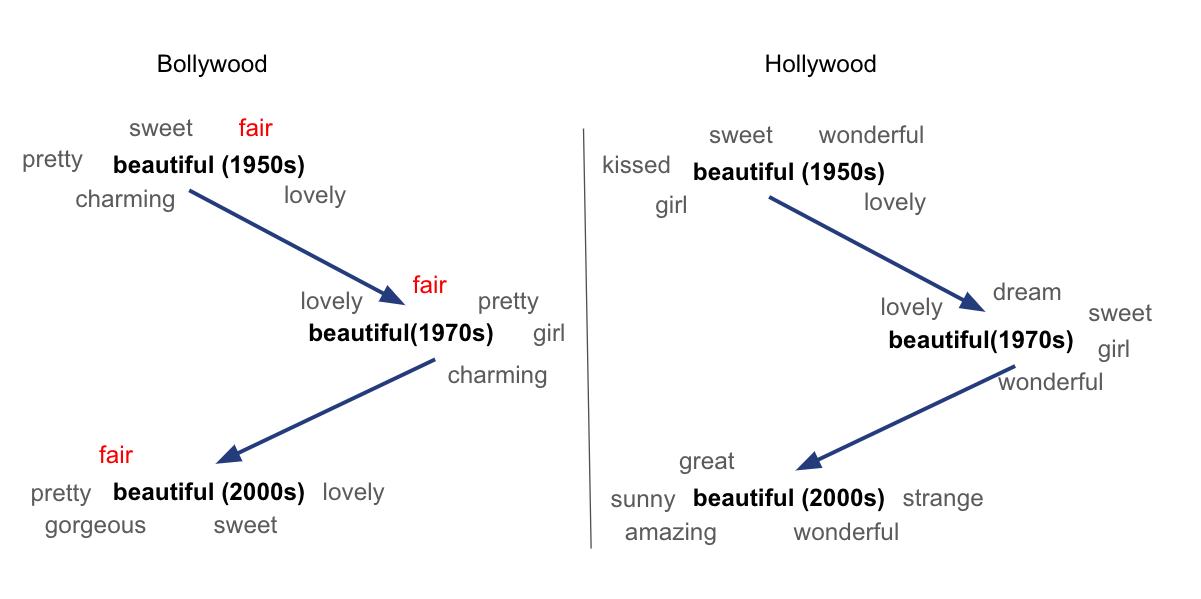}}
% \vspace{-0.3cm}
    \caption{Nearest neighbors of \texttt{beautiful} over the years}
    \label{fig:beautiful}
\end{figure}

We first present our cloze test results with the probe
"\emph{A beautiful woman should have} \texttt{[MASK]} \emph{skin.}" in Table~\ref{tab:allClozeSkin}. We note that while \texttt{BERT}$_\mathit{base}$ model predicts \emph{soft} in place of \texttt{[MASK]}, all fine-tuned \texttt{BERT} models on the film corpora predict \emph{\textcolor{red}{fair}} as the top choice. Figure~\ref{fig:beautiful} visualizes the nearest neighbors of \texttt{beautiful} in our aligned embedding spaces of Hollywood and Bollywood sub-corpora. 

%This showcases that films give an undue importance in associating beauty with fair skin, not only in older periods, but also in recent times. (Results in Appendix)
 
As shown in Figure~\ref{fig:beautiful} and Table~\ref{tab:allClozeSkin}, the age-old affinity toward lighter skin in Indian culture~\cite{dlova2015skin,chattopadhyay2019fair,madhukalya_2020} is reflected through the consistent presence of \texttt{fair} among the nearest neighbors of all three Bollywood sub-corpora. Although our cloze tests indicates Hollywood also exhibits bias towards lighter skin color, our diachronic word embedding analysis reveals that possibly the bias is less pronounced than in Bollywood.   \\

\section{Subtler Biases}

\emph{\textbf{RQ 4:} Does Bollywood reflect the well-documented son's preference in medical and social science research? How has the sentiment around retrograde social practices such as dowry evolved?} \\

Occupational stereotypes aside, 
a diachronic corpus may reveal subtler forms of biases. In our next analysis, we seek to analyze two seemingly disconnected aspects: son's preference and perception of dowry. 
Son's preference in India is a well-documented phenomenon and skewed sex ratio, female feticide and higher child mortality rate for girls have attracted policymakers' attention~\cite{pande2006preference,birth,nihcitation,birth2}. In order to prevent female feticide, in 1994, the Parliament of India enacted the Pre-Conception and Pre-Natal Diagnostic Techniques (PCPNDT) Act also known as the Prohibition of Sex Selection Act that effectively rendered prenatal sex discernment illegal.

Similarly, certain retrograde practice such as dowry, can influence son's preference as a girl-child might be looked upon as financial burden~\cite{diamond2008too}. The `dowry' system  has plagued the Indian society for a long time~\cite{dalmia2005institution}. Dowry refers to a transaction of tangible financial objects in the form of durable goods, cash, and real or movable property between the bride's family gives and the bridegroom, his parents and his relatives as a condition of the marriage. Although legally, dowry has been prohibited in India since 1961~\cite{rao1973dowry}, this practice has continued well after its legal prohibition and has a strong link to social crises such as female foeticide~\cite{ghansham2002female}, domestic abuse and violence~\cite{banerjee2014dowry,rastogi2006dowry}, and dowry deaths~\cite{ahmad2008dowry}. However, while the practice continued, recent studies have reported positive changes in the society where the general attitude towards the system has become negative~\cite{srinivasan2004dowry}.

\subsection{Son's preference}

A popular Bollywood plot point is the introduction of a child into the family. Approximately, every 1 in 10 collected movies had a scene involving birth of a child. 
We were curious to analyze when a child is born in a Bollywood movie, is it a boy or a girl? We retrieve the dialogues talking about childbirth using a template based approach, by searching for the following keywords -  \texttt{`birth',`baby',`pregnant',`pregnancy',`congratulations'} as well as phrases -  \texttt{``It's a boy'',``It's a girl''}. We annotated the retrieved dialogues related to childbirth, and perform a temporal analysis. 

Let $\mathcal{N}_\mathit{w}$ denote the number of times a dialogue talking about the baby's gender $w$ appears in a corpus. We define  \emph{Male Birth Ratio (MBR)} as follows:\\
    $\mathit{MBR} = \frac{\mathcal{N}_\mathit{boy}}{\mathcal{N}_\mathit{boy} + \mathcal{N}_\mathit{girl}}*100
$. Table \ref{tab:MBR} suggests that the family dynamics portrayed in Bollywood movies have shown considerable shift, with the $\mathit{MBR}$ being 73.9 in old movies, to almost achieving parity (54.5) in newer movies.\\

\begin{table}[h]
\centering
\begin{tabular}{|l|l|l|l|}
\hline
\Tstrut
                               & $\mathit{Old}$ & $\mathit{Mid}$ & $\mathit{New}$ \\ \hline
                \Tstrut
$\mathit{MBR}$         & 73.9         & 76.4 & 54.5        \\ \hline
\end{tabular}
\vspace{0.2cm}
\caption{The Male Birth Ratio (MBR) calculated based on Bollywood movie dialogues}
\label{tab:MBR}
\end{table}

 %This can be attributed partly due to the outcry on social media over the last few years, surrounding gender disparity in the entertainment industry \cite{bhattacharya}, as well as the positive progression taking place in India with regards to women rights. 

\begin{figure}[h]
\centerline{\includegraphics[frame,height=6cm,width=7.5cm ]{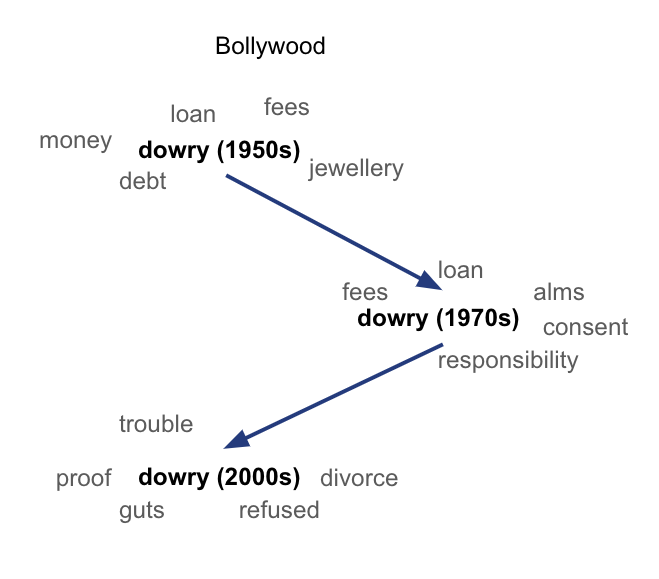}}
% \vspace{-0.3cm}
    \caption{Nearest neighbors of \texttt{dowry} over the years.}
    \label{fig:dowry}
\end{figure}

\subsection{Dowry}
 
As shown in Figure~\ref{fig:dowry}, we observe that while nouns such as \texttt{money}, \texttt{debt}, \texttt{jewellery}, \texttt{fees}, and \texttt{loan} are the nearest neighbors in older films indicating compliance to this practice, modern films exhibit non-compliance (e.g., \texttt{guts} and \texttt{refused}) and indicate some of the consequences of such non-compliance (e.g., \texttt{divorce} and  \texttt{trouble}) in the form of nearest neighbors.   
We find that indeed, films provide a snapshot of cultural values of a particular country thus allowing us  to gauge the progress of a nation over time.

% \subsection{Looking at gender bias through the lens of childbirth and dowry}

% \emph{\textbf{RQ 1} How is gender bias reflected through events of childbirth in a film?}

% \noindent \emph{\textbf{RQ 2} Harvard CRCS Can we track evolving trends of retrograde social practices such as dowry?}

% Our mixed method analyses take two very different paths to analyze these two aspects. We use alignment of diachronic word-embedding ... to analyze the first ...

\section{Broader Representation Questions}

\subsection{Georgraphic representation}

\emph{\textbf{RQ 5:} Which geographical areas have been consistently underrepresented in the Indian film industry?} \\

We now shift our focus to two broader representational aspects: geographical and religious representations. Of the 28 states in India, we next present an analysis of the relative representation of each of these states and major Indian cities. 

\begin{figure}%
\centering
\subfigure[1950--1999]{%
\label{fig:first}%
\includegraphics[height=1.5in]{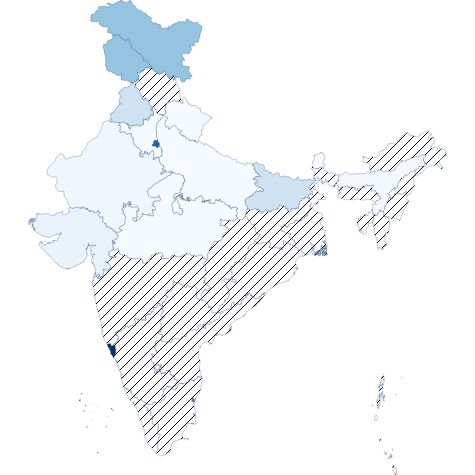}\label{fig:1950}}%
\qquad
\subfigure[2000--2020]{%
\label{fig:second}%
\includegraphics[height=1.5in]{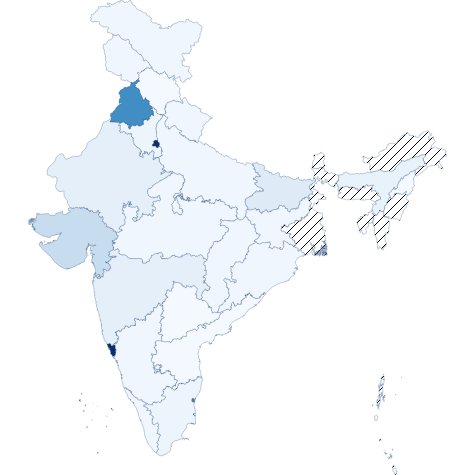}\label{fig:2000}}%
\qquad
\subfigure[No representation]{%
\label{fig:third}%
\includegraphics[height=1.5in]{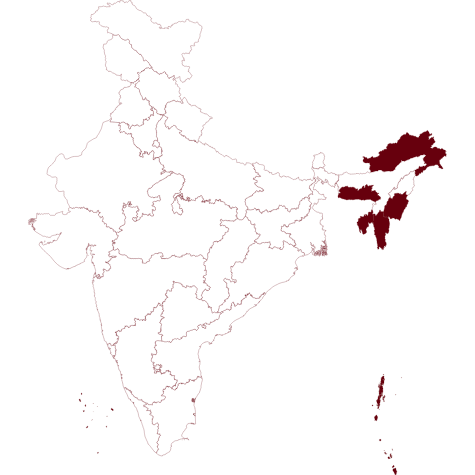}\label{fig:unused}}%
\caption{(a) Geographical Representation in films from 1950s to 2000s (b) Geographical Representation in films post 2000 (c) States never mentioned in our corpus in the last 70 years. The base maps used for this plot are sourced from the Government of India. The authors are aware that these maps include disputed territories. These maps do not constitute judgments on existing disputes.}
\label{fig:geographic_locations}
\end{figure}

From the beginning, Bollywood has it's roots in Mumbai, and Delhi is India's capital. Hence, it is not surprising that the cities are mentioned heavily across all time periods (see Table~\ref{tab:city_mentions}). Figure~\ref{fig:1950} and \ref{fig:2000} compare the geographic representations in the most recent 20 years with the rest of our corpus spanning 1950--2000. We observe that initially based out of major hotspots of Delhi, Goa and cities like Mumbai, recent Bollywood content is geographically more inclusive. However, A key point we highlight in Figure~\ref{fig:unused} is that, in line with prior research on underrepresentation of North Eastern states in news content \cite{haoginlen}, there is severe underrepresentation of Bollywood content surrounding these North Eastern states. There have been zero mentions of the states of Manipur, Arunachal Pradesh, Meghalaya, Tripura, Mizoram in over 500 movies across 70 years. 

%From the beginning, Bollywood has it's roots in Mumbai, the capital of Maharashtra in India. In this part of the analysis, we explore the movie content based on locations mentioned in the dialogues, to gain an insight into the geographical expansion of Bollywood in recent years. 

%\begin{figure}[htb]
%\centerline{\includegraphics[width=0.32\textwidth]{samples/unused_demographics.png}}
% \vspace{-0.3cm}
%    \caption{States never mentioned in Bollywood movies. The base maps used for this plot are sourced from the Government of India. The authors are aware that these maps include disputed territories. These maps do not constitute judgments on existing disputes.}
%    \label{fig:unusedstates}
%\end{figure}

%Table \ref{tab:city_mentions} gives us an indication that over the years, the cities mentioned in movies have changed significantly, and Mumbai, Delhi have remained the focal points irrespective of time periods. 

\begin{table}[]
\centering
\begin{tabular}{|l|l|l|}
\hline
\Tstrut
$\mathit{Old}$    & $\mathit{Mid}$      & $\mathit{New}$            \\ \hline
\Tstrut
Bombay/Mumbai {(}55{)} & Bombay/Mumbai {(}73{)} & Bombay/Mumbai {(}91{)}       \\ \hline
\Tstrut
Agra {(}50{)}          & Delhi {(}46{)}         & Delhi {(}63{)}               \\ \hline
\Tstrut
Delhi {(}30{)}         & Agra {(}46{)}          & Agra {(}42{)}                \\ \hline
\Tstrut
Daman {(}22{)}         & Daman {(}14{)}         & Bangalore/Bengaluru {(}11{)} \\ \hline
\Tstrut
Lucknow {(}14{)}       & Rampur {(}13{)}        & Pune {(}10{)}                \\ \hline
\Tstrut
Mathura {(}6{)}        & Lucknow {(}12{)}       & Hyderabad {(}9{)}            \\ \hline
\Tstrut
Srinagar {(}6{)}       & Pune {(}9{)}           & Amritsar {(}9{)}             \\ \hline
\Tstrut
Jalgaon {(}4{)}        & Bangalore {(}8{)}      & Lucknow {(}8{)}              \\ \hline
\Tstrut
Allahabad {(}4{)}      & Nagpur {(}7{)}         & Chandigarh {(}6{)}           \\ \hline

\end{tabular}
\vspace{0.5cm}
\caption{City mentions in movies}
\label{tab:city_mentions}
\end{table}

\subsection{Religious representation}

\emph{\textbf{RQ 6:} Can we gain an insight into the religious representation of a country, through a film corpus spanning 70 years?} \\

\begin{table}[h]
\centering
\begin{tabular}{|l|}
\hline
\Tstrut 
Most-frequent surnames \\ \hline

\Tstrut 
\begin{tabular}[c]{@{}l@{}}Singh, Krishna, Khan, Rai, Ali, Kapoor, Sharma, Mohan, Prasad, Khanna, Shah, \\ Lal, Thakur, Dev, Shekhar, Chaudhary, Gandhi, Verma, Gupta, Prakash, Rana, Nath, \\Patel, Pandey, Roy, Pandit, Saxena, Mathur, Roshan, Bachchan, Pal, Mehta, \\ Narayan, Das, Rode, Dayal, Mehra, Bhagat, Shastri, Chandra, Patil, Banerjee, \\ Tilak, Rao, Tripathi, Yadav, Kumari, Suman, Mukherjee, Bhatia, Acharya, \\ Chatterjee, Rehman, Iyer\end{tabular} \\ 
\hline

\end{tabular}
\vspace{0.2cm}
\caption{The top surnames occurring in Bollywood movies (in decreasing order of frequency)}
\label{tab:surnames}
\end{table}

\begin{table}[h]
\centering
\begin{tabular}{|l|}
\hline
\Tstrut
Surnames of doctors \\ \hline
\Tstrut
\begin{tabular}[c]{@{}l@{}}Kapoor, Chopra, Khurana, Tripathi, Kapoor, Ansari, Awasthi, \\Kothari, Mathur, Puri, Nayak, Bhalerao, Sawant, Tandon, Swamy, Banerjee, Verma, \\ Rana, Ruby, Singh, Shrivastav, Khanna, Bhandari, Tiwari, Saxena, Shinde, Mehta, \\ Goenka, Kumar, Goswami\end{tabular} \\ \hline
\end{tabular}
\vspace{0.2cm}
\caption{Surnames of doctors in Bollywood movies}
\label{tab:doc_surnames}
\end{table}

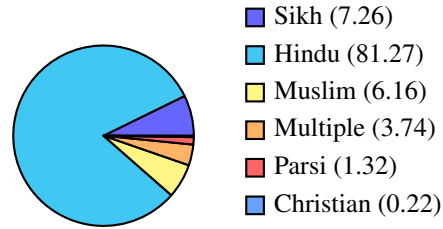
\begin{figure}[h!]
    \centering
    \begin{tikzpicture}[scale=0.40]
  \pie[text=legend,hide number]{7.26/Sikh (7.26),81.27/Hindu (81.27),6.16/Muslim (6.16),3.74/Multiple (3.74),1.32/Parsi (1.32),0.22/Christian (0.22)}
  \end{tikzpicture}
    \caption{Religious Representation in Old movies}
    \label{fig:old_religion}
\end{figure}

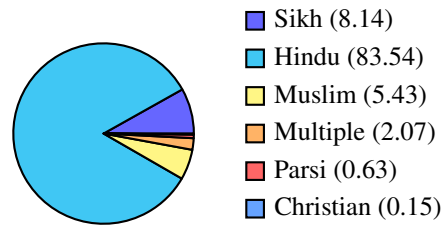
\begin{figure}[h!]
    \centering
    \begin{tikzpicture}[scale=0.40]
  \pie[text=legend,hide number]{8.14/Sikh (8.14),83.54/Hindu (83.54),5.43/Muslim (5.43),2.07/Multiple (2.07),0.63/Parsi (0.63),0.15/Christian (0.15)}
\end{tikzpicture}
    \caption{Religious Representation in Mid movies}
    \label{fig:mid_religion}
\end{figure}

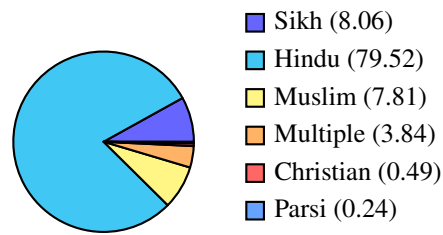
\begin{figure}[h!]
    \centering
    \begin{tikzpicture}[scale=0.40]
  \pie[text=legend,hide number]{8.06/Sikh (8.06),79.52/Hindu (79.52),7.81/Muslim (7.81),3.84/Multiple (3.84),0.49/Christian (0.49),
0.24/Parsi (0.24)}
\end{tikzpicture}
    \caption{Religious Representation in New movies}
    \label{fig:new_religion}
\end{figure}

India is a diverse country with 6 main religions, 22 major languages and more than 700 dialects. According to the 2011 census \cite{religion_census}, the religious distribution among the Indian population is 79.8\% follow Hinduism, 14.2\% adheres to Islam, 1.72\% adheres to Sikhism, 2.3\% adheres to Christianity, 0.7\% adheres to Buddhism and 0.37\% adheres to Jainism. There has been discussion in the past, regarding the religious stereotypes portrayed in Bollywood \cite{haris}, and we build upon this by understanding the religious representation present in the large corpus of Bollywood subtitles. The surnames appearing in the movies (e.g., Mrs. Kapoor, Mr. Khan etc.) are annotated manually by two annotators, with each surname given one label from the list of labels: Hindu; Muslim; Sikh; Christian; Parsi; and Multiple. The annotators achieved a Cohen $\kappa$ score of 0.8879 indicating high inter-rater agreement. The discrepancies were resolved by the annotators through a follow-up adjudication process and by consulting relevant literature. 
Figures \ref{fig:old_religion}, \ref{fig:mid_religion}, \ref{fig:new_religion} provide the religion distribution obtained in movies. We note that: (1) the distribution is more or less consistent with the census numbers; (2) representation for other religions has increased in recent years; and (3) the representation of Muslims is slightly less than the community's population share. 

% Using the \centering command instead of \begin{center} ... \end{center} will save space
% Positioning your figure at the top of the page will save space and make the paper more readable
% Using 0.95\columnwidth in conjunction with the

Table \ref{tab:doc_surnames} indicates the surnames of the doctors occuring in Bollywood movies. To retrieve these surnames from the subtitles, we employ a template based approach, searching for keywords like \texttt{`Dr.', `doctor'} from our corpus. While a broader religious representation is observed in our overall results, the observed representation for the medical profession is quite skewed, with large number of surnames being Brahmins (the uppermost caste in the Hindu caste-system in India).

\section{Economic Signals}

\emph{\textbf{RQ 7:} Can we extract economic signals through popular film dialogues?} \\

\begin{table}[h!]
\centering
\begin{tabular}{|l|l|l|l|}
\hline
\Tstrut
                               & $\mathit{Old}$ & $\mathit{Mid}$ & $\mathit{New}$ \\ \hline
                               \Tstrut
Predicted Average Ransom Amount & 594,805            & 10,959,940         & 29,688,280         \\ \hline
\Tstrut
Inflation adjusted & -            & 2,194,830         & 21,000,280         \\ \hline
\end{tabular}
\vspace{0.5cm}
\caption{Average amount for text completion results on the input sentence \emph{The ransom amount is} using fine-tuned \texttt{GPT-2} models. The inflation-adjusted values for 594,805 INR in 1960 is presented in bottom row.}
\label{tab:ransom}
\end{table}

By looking at a popular entertainment corpus of a developing nation, we are able to showcase the evolution of gender bias, evolving attitude towards social evils, geographic and religious representations. Can we detect economic signals as well, from the Bollywood dialogues? 100 rupees in 1958 are equivalent to 8,117.22 rupees in 2020 \cite{inr}. We seek to understand whether language models can capture these noisy signals. 
\texttt{GPT-2} \cite{gpt2}, a popular language model with more than 100 million parameters, has achieved state of the art results for text completion, zero shot transfer learning etc. \texttt{GPT-2} has been widely used for generating free form text, to create artificial newsletters, poems \cite{gwern_2019} etc. We noticed that the most common dialogues expressing monetary figures or large amounts of money were generally associated with ransom. For example, a sample dialogue is ``We have kidnapped your kid, the ransom amount is 2 million Rupees.'' %A ransom amount of 1 lakh Rupees (100,000 INR) would be an exorbitant amount in the 1950s. %On the other hand, 50 million Rupees (50,000,000 INR) can be considered expensive in today's times. This shows us that the purchasing power of money has changed drastically over the last 70 years. 

To understand this change at a large scale, we fine-tune \texttt{GPT-2}  on three Bollywood sub-corpora, each belonging to films from different time periods, for the end goal of text/dialogue generation. On these finetuned models, we input the sentence \emph{``The ransom amount is''} and analyze the generated text by the model. Table \ref{tab:ransom} showcases the average amount across 100 generated samples from the finetuned models. We note that our while our predicted values  overestimate the inflation rate,  the ransom amounts capture the general increasing pattern and have increased significantly over time. \\

\section{Evolving National Priorities as Reflected in Movies and Responsible Censorship}

\emph{\textbf{RQ 8:} How are religions perceived in movies? Can we track evolving national priorities from popular entertainment?}

\subsection{Responsible censorship}
\begin{figure}[htb]
\centerline{\includegraphics[frame,totalheight=5cm,width=0.8\linewidth]{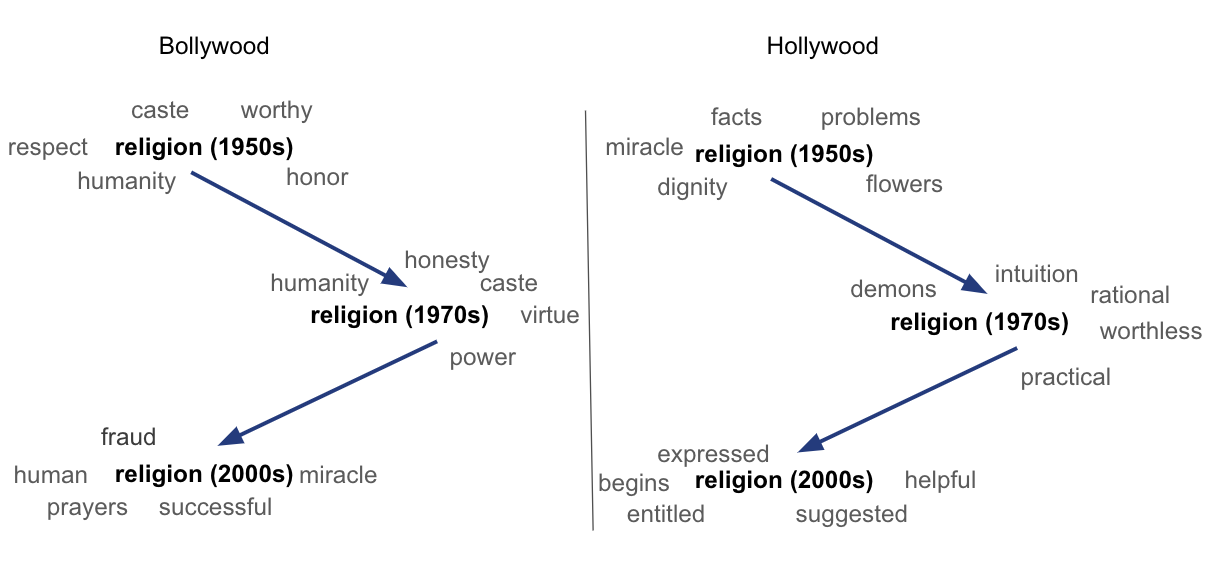}}
% \vspace{-0.3cm}
    \caption{\texttt{religion} over the years}
    \label{fig:religion}
\end{figure}

India has faced two major partitions over the last 70 years, which have resulted in considerable religious turmoil and riots \cite{riots}. The Censor Board is a governing body which along with giving each movie a certification, has the ability to remove offensive or controversial content or in some extreme cases, completely ban films from being screened in  theatres. With religion being a contentious topic in India, offensive terms surrounding it are also discouraged, and this has been constant throughout the years. To validate this hypothesis, we look at the word `religion' and how it has evolved over the years. Figure \ref{fig:religion} indicates that `religion' is always accompanied with neutral/mild terms, and movie dialogues have stayed away from using extreme or hateful terms surrounding religion. 

\begin{figure}[htb]
% \vspace{-.15cm}
\centering
\subfigure[\texttt{Hindu} over the years]{%
\includegraphics[frame,width = 0.48 \textwidth, height=0.35\textwidth]{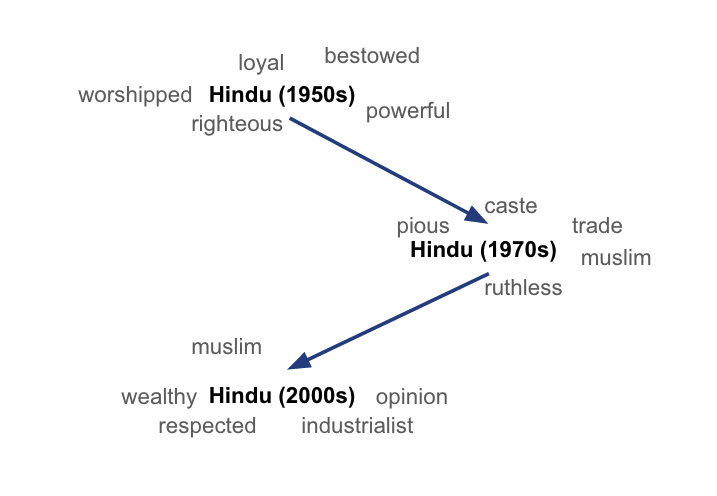}
\label{fig:Hindu_transition}}
\subfigure[\texttt{Muslim} over the years]{%
\includegraphics[frame,width = 0.48 \textwidth, height=0.35\textwidth]{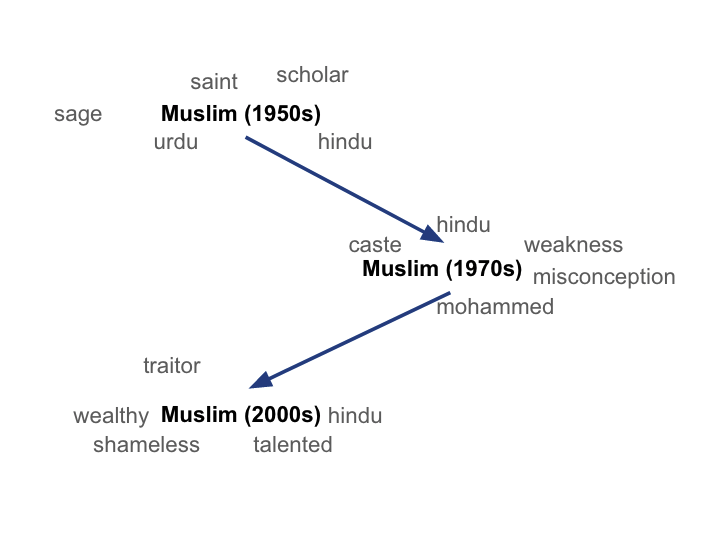}
 \label{fig:Muslim_transition}}
\caption{Nearest neighbors of \texttt{Hindu} and \texttt{Muslim} over the years.}
\end{figure}

Along with understanding the nearest neighbors of `religion', we wanted to understand the discussion surrounding the two biggest religions in India, `Hindu' and `Muslim' (Figures \ref{fig:Hindu_transition}, \ref{fig:Muslim_transition}). We contrast our findings on the movie corpus with prior research involving raw social media data \cite{ECAIElection}. While negative words like \texttt{ruthless}, \texttt{shameless}, and \texttt{traitor} creeping up in newer movies might indicate religious polarization, we notice that words like \texttt{terrorists} found in social media data  \cite{ECAIElection}, do not surface among the nearest neighbors. Along with word embedding analysis, we analyze the \texttt{BERT} cloze tests for the probes (1) \emph{Hindus are }\texttt{[MASK]} and (2) \emph{Muslims are }\texttt{[MASK]}. For both probes, we do not notice completions such as \texttt{terrorists} or \texttt{fools} previously reported in \cite{ECAIElection}. This indicates that while recent social media analyses might indicate religious polarization, film certification board largely ensured movie content do not reflect such  divide. % the difference between text-in-the-wild vs curated text through responsible film censorship, performed by the Central Board of Film Certification (CBFC) in India.  

\begin{table*}[h]

{
\scriptsize
\begin{center}
     \begin{tabular}{| p{4.1cm} | p{4.1cm} | p{4.1cm} |}
    \hline
 \textbf{\texttt{BERT}\textsubscript{base}}                                                               & \textbf{\texttt{BERT}\textsubscript{$\mathcal{D}_\mathit{bolly}^\mathit{old}$}}                                                & \textbf{\texttt{BERT}\textsubscript{$\mathcal{D}_\mathit{bolly}^\mathit{new}$}} \\ \hline
    corruption, poverty, malaria, pollution, hunger, terrorism, unemployment, drought, famine, war, tourism & poverty, love, war, hunger, unemployment, india, famine, money, marriage, education, \textcolor{red}{kashmir} & poverty, \textcolor{red}{pakistan}, \textcolor{red}{kashmir}, terrorism, corruption, india, drugs, dowry, unemployment, hunger, rape \\ \hline

    \end{tabular}
\end{center}
\caption{{Cloze test results for \emph{The biggest problem in India is} \texttt{[MASK]}.}}
\label{tab:cloze4}}
%%\vspace{-.2in}
\end{table*}

\begin{table*}[h]

{
\scriptsize
\begin{center}
     \begin{tabular}{| p{4.1cm} | p{4.1cm} | p{4.1cm} |}
    \hline
 \textbf{\texttt{BERT}\textsubscript{base}}                                                               & \textbf{\texttt{BERT}\textsubscript{$\mathcal{D}_\mathit{holly}^\mathit{old}$}}                                                & \textbf{\texttt{BERT}\textsubscript{$\mathcal{D}_\mathit{holly}^\mathit{new}$}} \\ \hline
 poverty, corruption, unemployment, crime, terrorism, racism, pollution, hunger, war, cancer, inequality & war, poverty, money, unemployment, slavery, immigration, alcoholism, education, imperialism, \textcolor{red}{russia}, hunger & poverty, slavery, immigration, unemployment, money, war, \textcolor{red}{racism}, hunger, communism, america, education \\ \hline

    \end{tabular}
\end{center}
\caption{{Cloze test results for \emph{The biggest problem in America is} \texttt{[MASK]}.}}
\label{tab:cloze5}}
%%\vspace{-.2in}
\end{table*}

\subsection{Evolving National Priorities as Reflected in Movies}
Similar to \texttt{BERT}'s cloze test applications to uncover gender and racial bias showcased in earlier sections, we employ \texttt{BERT} to analyze evolving national priorities using the following two cloze tests. 
\begin{enumerate}
    \item For Bollywood: \emph{The biggest problem in India is} \texttt{[MASK]}. 
    \item For Hollywood: \emph{The biggest problem in America is} \texttt{[MASK]}. 
\end{enumerate}

We observe that the  dynamic political conditions are reflected in the completion results in Tables \ref{tab:cloze4} and \ref{tab:cloze5} (e.g., \texttt{Kashmir}, \texttt{Pakistan} and \texttt{Russia})~\cite{kashmirbook,kashmirbook2,lafeber1972america}. We also note that the list of ongoing problems in the U.S. contains the major issue on which the 2020 election was fought over: racism.

\section{Future Directions and Conclusion}

\subsection{Future directions}

While our current work focused on answering research questions related to inclusion and bias, our corpus may find use in several other interesting applications. Here, we outline an exploratory analysis on computational creativity. 

In recent years, the emergence of neural architectures and language models like \texttt{GPT-2} \cite{gpt2} with millions of parameters have resulted in rapid advancements for various NLP tasks. These models has proved efficient in generating artificial poems \cite{poem1,poem2}, stories, news articles with just a few epochs of fine-tuning. We employ a similar strategy to generate dialogues based on the subtitles collected from more than 2000 films, using \texttt{GPT-2} \cite{gpt2}. To the best of our knowledge, ours is the first ever attempt to automatically generate artificial Bollywood dialogues with human evaluations conducted by annotators with professional  experience with the  Bollywood industry.  

\begin{table}[h]
\centering
\begin{tabular}{|p{2.3cm}|p{10.7cm}|}
\hline
\textbf{Input Probe}                                  & \textbf{Artificially Generated Subtitles}                                                                                                                                                                                                                                                                                                     \\ \hline
The ransom for kidnapping                             & \begin{tabular}[c]{@{}l@{}}The ransom for kidnapping hasn't been paid yet.\\ We'll ask for the ransom at the next hearing.\\ Then call up Mr. Trivedi and ask him for the ransom at the next hearing.\\ Fine.\\ Then we'll ask for the next hearing at our doorstep.\\ And he'll refuse.\\ Yes, tell me.\\ Come.\\ Sir, please listen to me.\end{tabular} \\ \hline
Your eyes are so                                      & \begin{tabular}[c]{@{}l@{}}Your eyes are so beautiful.\\ They don't let me sleep."\\ They are just dreams."\\ My heart is singing."\\ Don't let it go, beloved."\\ Don't let it go, beloved."\end{tabular}                                                                                                                                                \\ \hline
What does the police have to say about this incident? & \begin{tabular}[c]{@{}l@{}}What does the police have to say about this incident?\\ What do you mean?\\ They don't have a witness, madam.\\ No case can be filed under section 302.\\ They only record the statement of the complainants.\\ The police is much more capable than this.\\ Their power and reach is much more impressive.\end{tabular}       \\ \hline
\end{tabular}
\vspace{0.5cm}
\caption{Machine generated subtitles using the GPT2 model. We provide the input probe to the model and output the free form text completion generated by the finetuned LM on our corpus.}
\label{tab:artificial_scripts}
\end{table}

\textbf{Human Evaluation of the generated dialogues:} We employed two annotators either working professionally or closely associated with the Bollywood film industry, to guess the dialogues into one of the two labels, Real or Generated. We provided a list of 5 real dialogue snippets taken from movies, along with 6 artificially generated subtitles. $\mathit{Annotator}_1$ labelled 4 out of 6 artificial dialogues as being real, while $\mathit{annotator}_2$ labelled 3 out of 6 artificial dialogues as being real, thus showcasing the efficacy of recent advancement in language models in generating human-like movie diaolgues. 

\subsection{Conclusions}
In this paper, we analyzed how social biases and subtle gender biases get reflected on diachronic corpora of popular entertainment. Our research indicates that our NLP methods are capable of uncovering important social signals. Our results demonstrate that indeed, societal changes do get reflected in popular content. But does popular entertainment also influence the society in turn? A recent movie on acid attack, Chhapak, was inspired from a true story of an acid attack survivor who set up an NGO and was a recipient of the \emph{International Women of Courage} \cite{firstpost_2014} award. Her biopic and her initiative of \emph{Stop Acid Sale} when released, triggered regulatory legislation that made it difficult to buy certain types of acids without legal authorization. Devising NLP methods to identify how popular entertainment influences society will be a worthy future research challenge. 

% Our future lines of work include (1) tracking moral sentiment~\cite{xie-etal-2019-text}; (2) exploring debiasing techniques~\cite{manzini-etal-2019-black}; (3) analyzing representation of other minorities (e.g., Muslims, dalits and the LGBTQ community); and (4) exploring a newly-proposed machine-translation data set distance measure~\cite{khudabukhsh2020dont} to uncover contrasting social aspects.  

% While our current focus is on characterizing gender bias and representation, our rich data set has untapped potential in analyzing a wide range of research questions. For instance, in line with the cloze tests conducted in~\cite{ECAIElection}, Table~\ref{tab:cloze4} and \ref{tab:cloze5} list the results for two additional cloze tests: (1) \emph{The biggest problem of India is} \texttt{[MASK]}. (referred as $\mathit{cloze}_4$) - For Bollywood and (2) \emph{The biggest problem of America is} \texttt{[MASK]}. (referred as $\mathit{cloze}_5$) - For Hollywood. 

% We observe that the  dynamic political conditions are reflected in the completion results (e.g., \texttt{Kashmir}, \texttt{Pakistan} and \texttt{Russia})~\cite{kashmirbook,kashmirbook2,lafeber1972america}. We also note that the list of ongoing problems in the U.S. contains the major issue on which the 2020 election is being fought over: racism.

% Future Work - 
% 1. Computational Creativity
% 2. Answering Causal Questions

% Conclusions: summarize our findings.


\clearpage
\bibliographystyle{unsrt}

%\bibliography{references.bib}

\appendix
\newpage
\section{APPENDIX}
\subsection{Implementation Details}
Experiments are conducted on a Google Colab Pro instance, using the Tesla V100 and P100 GPUs provided by Google in the Colab notebook. 

We follow the standard preprocessing steps recommended to fine-tune BERT language model. For our task, we use the bert-base-uncased pretrained English model, with the following parameter details: 12 transformer layers, hidden state length of 768, 12 attention heads, 110M overall parameters. The pre-trained model is fine tuned on the target corpus using the training parameters showcased below. \begin{itemize}[itemsep=0.4mm, parsep=0pt]
\item Batch size: 16
\item Maximum sequence length: 128
\item Maximum predictions per sequence: 20
\item Fine-tuning steps: 10,000
\item Warmup steps: 10
\item Learning rate: 2e-5
\end{itemize}

For fine-tuning the language model for free form text completion tasks, we use the smallest GPT2 model with 124M parameters, trained for 10,000 steps. The results showcased in Table \ref{tab:artificial_scripts} are obtained from the fine-tuned model by setting length=250 and temperature=0.9.

\subsection{BERT Fine Tuning Results}

\begin{table*}[h]
\scriptsize
\begin{center}
     \begin{tabular}{| p{1cm}  | p{2.1cm} | p{2.1cm} | p{2.1cm} | p{2.1cm} | p{2.1cm} |}
    \hline
\textbf{Probe} & \textbf{\texttt{BERT}\textsubscript{base}}                                                               & \textbf{\texttt{BERT}\textsubscript{$\mathcal{D}_\mathit{bolly}^\mathit{old}$}}                                                & \textbf{\texttt{BERT}\textsubscript{$\mathcal{D}_\mathit{bolly}^\mathit{new}$}}                                                       & \textbf{\texttt{BERT}\textsubscript{$\mathcal{D}_\mathit{holly}^\mathit{old}$}}                                                            & \textbf{\texttt{BERT}\textsubscript{$\mathcal{D}_\mathit{holly}^\mathit{new}$}}                                                              \\ \hline
$\mathit{cloze}_1$         & man, widow, woman, \textcolor{blue}{doctor}, \textcolor{red}{slave}, \textcolor{blue}{soldier}, bachelor, \textcolor{blue}{merchant}, \textcolor{blue}{farmer}, \textcolor{blue}{lawyer}, \textcolor{red}{servant} [\textcolor{black}{\textbf{4.8}}] & \textcolor{red}{prostitute}, \textcolor{red}{servant}, woman, \textcolor{red}{slave}, bachelor, \textcolor{blue}{doctor}, \textcolor{blue}{lawyer}, man, widow, \textcolor{red}{maid}, \textcolor{blue}{worker} [\textcolor{black}{\textbf{4.64}}] & \textcolor{blue}{doctor}, woman, \textcolor{red}{servant}, \textcolor{blue}{lawyer}, \textcolor{red}{maid}, hindu, \textcolor{blue}{nurse}, \textcolor{blue}{teacher}, \textcolor{blue}{gardener}, lady, man [\textcolor{black}{\textbf{5.7}}] & woman, \textcolor{red}{slave}, \textcolor{red}{servant}, \textcolor{blue}{nurse}, lady, man, \textcolor{blue}{teacher}, \textcolor{blue}{lawyer}, peasant, \textcolor{red}{maid}, wife [\textcolor{black}{\textbf{5.3}}] & woman, \textcolor{blue}{lawyer}, \textcolor{blue}{doctor}, \textcolor{blue}{nurse}, \textcolor{blue}{teacher}, man, \textcolor{blue}{writer}, \textcolor{blue}{secretary}, \textcolor{red}{prostitute}, professional, \textcolor{blue}{carpenter} [\textcolor{black}{\textbf{5.7}}] \\ \hline
$\mathit{cloze}_2$         & man, \textcolor{blue}{soldier}, gentleman, \textcolor{blue}{farmer}, \textcolor{blue}{merchant}, woman, \textcolor{red}{slave}, bachelor, \textcolor{blue}{doctor}, \textcolor{blue}{carpenter}, \textcolor{red}{servant} [\textcolor{black}{\textbf{5.48}}] & man, \textcolor{blue}{gentleman}, \textcolor{blue}{lawyer}, lawyer, \textcolor{red}{servant}, \textcolor{blue}{doctor}, \textcolor{blue}{farmer}, \textcolor{blue}{worker}, \textcolor{blue}{craftsman}, \textcolor{red}{slave}, \textcolor{red}{criminal} [\textcolor{black}{\textbf{5.0}}] & \textcolor{blue}{doctor}, \textcolor{blue}{lawyer}, \textcolor{blue}{policeman}, man, \textcolor{blue}{farmer}, bachelor, \textcolor{blue}{gardener}, \textcolor{red}{servant}, \textcolor{blue}{soldier}, \textcolor{blue}{mechanic}, \textcolor{blue}{builder} [\textcolor{black}{\textbf{5.3}}] & \textcolor{blue}{carpenter}, \textcolor{blue}{policeman}, \textcolor{blue}{lawyer}, \textcolor{blue}{soldier}, \textcolor{blue}{farmer}, \textcolor{blue}{gentleman}, \textcolor{red}{servant}, man, peasant, \textcolor{red}{slave}, \textcolor{blue}{doctor} [\textcolor{black}{\textbf{5.0}}] & man, \textcolor{blue}{lawyer}, \textcolor{blue}{soldier}, \textcolor{blue}{doctor}, \textcolor{blue}{carpenter}, \textcolor{blue}{gentleman}, \textcolor{blue}{clergyman}, \textcolor{blue}{farmer}, \textcolor{blue}{writer}, \textcolor{blue}{craftsman}, \textcolor{blue}{minister} [\textcolor{black}{\textbf{5.78}}] \\ \hline
$\mathit{cloze}_3$         & soft, beautiful, pale, tanned, smooth & \textcolor{red}{fair}, no, pale, tanned, tan & \textcolor{red}{fair}, tanned, golden, smooth, pale & \textcolor{red}{fair}, pale, blue, golden, gold & \textcolor{red}{fair}, pale, tanned, golden, dark \\ \hline
    \end{tabular}
\end{center}
\vspace{0.5cm}
\caption{{Cloze test results. Predicted tokens are ranked by decreasing probability. Positive and negative words are color coded with blue and red, respectively. The number in the bracket represents the average valence score (obtained from~\protect\cite{warriner2013norms}) calculated for the answers to the cloze test.}}
\label{tab:allCloze}
%%\vspace{-.2in}
\end{table*}

\end{document}